\documentclass{ws-ijmpcs}

\begin{document}

\markboth{ARKA \& KIRK}{Lineraly polarized superluminal waves in pulsar winds}

%
\catchline{}{}{}{}{}
%

\title{Linearly polarized superluminal waves in pulsar winds
}

\author{IOANNA ARKA}

\address{Max Planck Institut f\"ur Kernphysik, Saupfercheckweg 1\\
Heidelberg, 69117, Germany\\ and \\
Institute de Plan\'etologie et d'Astrophysique de Grenoble, UMR 5274 \\ 
BP 53 F-38041 GRENOBLE \\
ioanna.arka@ujf-grenoble.fr}

\author{JOHN G. KIRK}

\address{Max Planck Institut f\"ur Kernphysik, Saupfercheckweg 1\\
Heidelberg, 69117, Germany\\
john.kirk@mpi-hd.mpg.de}

\maketitle

\begin{history}
\received{Day Month Year}
\revised{Day Month Year}
\end{history}

\begin{abstract}
Pulsar winds are the ideal environment for the study of non-linear 
electromagnetic waves. It is generally thought that a pulsar launches a striped 
wind, a magnetohydrodynamic entropy wave, where plasma sheets carried 
along with the flow separate regions of alternating magnetic field. But when 
the density drops below a critical value, or equivalently for distances from the 
pulsar greater than a critical radius, a strong superluminal wave can also 
propagate. In this contribution we discuss the conversion of the equatorial 
striped wind into a linearly polarized superluminal wave, and we argue that this mode is 
important for the conversion of Poynting flux to kinetic energy flux before 
the outflow reaches the termination shock.

\keywords{plasmas; waves; acceleration of particles; pulsars:general; stars: winds, outflows.}
\end{abstract}

\ccode{PACS numbers: 11.25.Hf, 123.1K}

\section{Introduction}	

Pulsar winds are relativistic outflows, consisting mainly of electrons and 
positrons. They are thought to be Poynting-flux dominated at launch, as quantified
by the magnetization parameter $\sigma$ which is the ratio of Poynting
to kinetic energy flux, with $\sigma \gg 1$. Far away from the light cylinder, 
which is at radius $r_{\rm LC} 
= c/\omega$ with $\omega$ the pulsar's rotational frequency, 
the magnetic field is predominantly toroidal with an amplitude falling as 
$1/r$ and reverses sign across a corrugated current sheet that separates
the two magnetic hemispheres. The wind propagates radially as a transverse
subluminal wave with speed equal to the bulk speed of the outflow, or 
equivalently Lorentz factor $\Gamma$. The opening angle around the equator 
of the region of alternating magnetic field depends on the misalignment of 
the magnetic and rotational axes of the pulsar, denoted by $\alpha$.

In the absence of dissipation, the wave propagates outwards from the light cylinder
while remaining Poynting-dominated\cite{kirkmochol}. However, 
after crossing the termination shock, the outflow's magnetization is low\cite{reesgunn}.
Where and when conversion of Poynting flux to particle energy
happens is still not clear. Magnetic reconnection in the current sheets
separating regions of opposite magnetic field polarity in the wind is a possible
dissipation mechanism, 
however this process is much 
too slow to account for the conversion in isolated pulsars like the Crab,
unless the pair injection by the pulsar is much larger than generally 
assumed\cite{coroniti}\cdash\cite{kirkskjaeraasen}. Another possibility is driven reconnection at the termination 
shock, which has been demonstrated to dissipate the striped wind's 
alternating field efficiently in some cases. This, however, requires even higher
particle injection rates than reconnection in the current sheets\cite{lyubarsky}\cdash\cite{sironispitkovsky}.

A solution can be provided in the form of non-linear superluminal waves,
i.e. electromagnetic waves of large amplitude with phase speed $v_{\phi} > c$. 
These can propagate outside of a critical radius $r_{\rm cr}$, at regions of 
lower particle density\cite{melatosmelrose} (as a contrast to the magnetohydrodynamic wind
which needs a minimum particle density). The conversion of the striped
wind to a superluminal mode is possible if $r_{\rm cr}$ is smaller than the
shock radius.
 
In the following we will investigate non-linear superluminal waves of linear polarization, which are relevant
in a region of opening angle $\alpha$ around the equator of a pulsar wind.
We identify the superluminal modes into which a striped wind can convert 
by imposing "jump conditions" across the (thin) region where conversion
happens. We will assume throughout that the waves can be considered as 
locally plane and purely transverse. These are excellent approximations far away from the light
cylinder
(at $\rho = r/r_{\rm LC} \gg 1$). The radial direction, which is the direction
of propagation of the wave, can be  identified with the $x-$direction in a 
cartesian system of coordinates. The surviving field components are then 
$E_y$ and $B_z$, corresponding to the azimuthal magnetic field and the polar electric field. 

\section{Transverse superluminal waves: the two-fluid approach}

The equations governing the evolution of a cold, two-component plasma are the continuity
equation and the equations of motion for each species, along with Maxwell's 
equations. We are looking for periodic solutions in a special frame of reference that
propagates with velocity $c^2/v_{\phi}$ with respect to the laboratory frame (in the laboratory frame, which coincides
with the pulsar frame, the wave phase speed is $v_{\phi}$). We will
call this special frame \lq\lq homogeneous \rq\rq or \lq\lq  H-frame\rq\rq. The 
advantage of this formulation is that in the H-frame all space dependence 
vanishes and only time dependence remains, and the phase is just $\omega_0 t$ 
with $\omega_0$ the wave frequency\cite{clemmow}. From now on quantities in the H-frame will be unprimed,
while quantities in the laboratory frame will be primed. 

In the H-frame, the conservation of magnetic
flux ($\mathbf{\nabla}\cdot \mathbf{B} = 0$) is automatic, and it can be seen immediately 
from Faraday's law that the magnetic field is constant. From Coulomb's law and the continuity equation we get $n_+\gamma_+ = 
n_-\gamma_-$, where
a plus(minus) subscript denotes positrons(electrons). Furthermore, the 
Lorentz factors are equal $\gamma_+ = \gamma_-$ and the other components
of the dimensionless four-velocity of the two species are related by:
\begin{eqnarray}
u_{x+} \,=\, u_{x-} \quad u_{y+} \,=\, -u_{y-} \quad u_{z+} \,=\, -u_{z-}
\enspace.
\end{eqnarray}Because of these equations we can drop the subscripts and 
solve the system in terms of the positron fluid quantities. It can be shown\cite{kennelpellat}
that all quantities can be expressed in terms of the electric field normalized to its largest amplitude
$y = E/E_0$:
\begin{eqnarray}
\gamma = \gamma_0 + 2 \gamma_0 (1-y^2)/q \\
u_x = \sqrt{\gamma_0^2-1} + 4\gamma_0 \lambda(1-y)/q \\
u_y = \pm \sqrt{\gamma^2-u_x^2-1} \\
u_z = 0 \,.
\end{eqnarray}In the above, quantities with a zero subscript express the initial condition at zero phase, and the field reaches
its largest amplitude at phase zero, $y_0=1$. Also, $\lambda = 
B/E_0$ is the ratio of the (constant) magnetic field to the electric field
amplitude, and $q$ is double the ratio at phase zero of the energy density in the particles to that in the electric
field:
\begin{equation}
q = \frac{32 \pi m c^2 \gamma_0^2 n_0}{E_0^2}.
\end{equation}
The electric field $y$ can be computed from a first-order differential equation, which in the general case ($\lambda\neq 0$) 
has no solution in 
closed form and has to be numerically integrated. The phase-averaged values of the above quantities, then, can
also be calculated numerically (for details see Ref.~\refcite{arkakirk}).

\section{Conserved quantities}

There are four conserved quantities in a pulsar wind, in the plane wave approximation. These are the phase-averaged
$x-$components of the particle, energy, momentum and magnetic flux densities. We will denote the
particle flux by $\left< J' \right>$, while the energy and momentum flux 
densities are the $\left< T'^{01} \right>$ and $\left<T'^{11}\right>$ components of the combined stress-energy 
four-tensor of fields 
and particles. Finally, the magnetic flux density can be replaced (from Faraday's law) by the phase-averaged 
electric field, so that the last conserved quantity can be written as $\left< E' \right>^2/(4\pi)$.

\begin{figure}[pt]
\centerline{\psfig{file=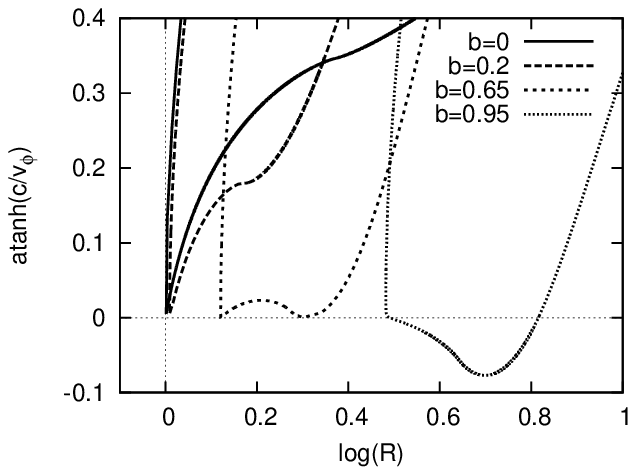,width=6.5cm}\psfig{file=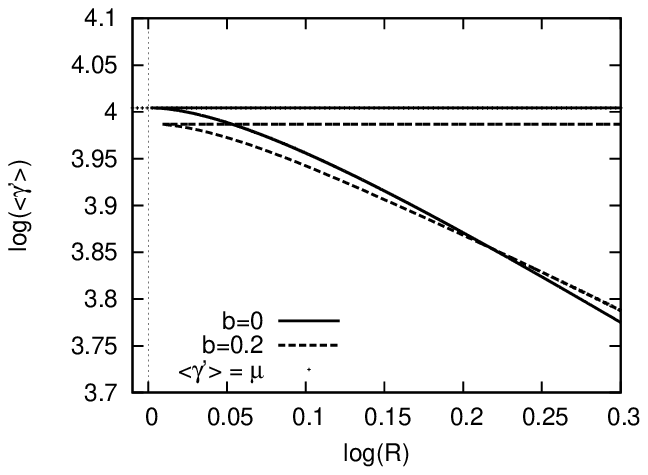,width=6.5cm}}
\vspace*{8pt}
\caption{On the left the index of refraction is plotted as a function of dimensionless radius $R$, for four different
cases of $b$. On the right, the phase-averaged Lorentz factor of the particles is plotted. We see that for low $b$, $
\left< \gamma'\right>\simeq \mu$, which means that almost the whole energy of the wind is now carried by the 
particle component of the flow.  \label{f1}}
\end{figure}

From these four quantitites we can construct three dimensionless ones, defined as
\begin{equation}
\mu = \frac{\left< T'^{01}\right>}{mc\left< J'\right>} \label{mugeneral} \end{equation}
\begin{equation}
\nu = \frac{\left< T'^{11}\right>}{mc\left< J'\right>}  
\end{equation}\begin{equation}
\eta = \frac{\left< E'\right>^2}{4\pi mc\left< J'\right>} \label{etageneral}\,.
\end{equation}

The first of these, $\mu$, is the energy per unit mass that each particle would have, if the entire energy flux
were carried by the particles. The three quantities (\ref{mugeneral})-(\ref{etageneral}) are conserved across a layer that marks 
the transition between the striped wind and the superluminal wave, and therefore they have to be expressed in terms of the 
quantities characterizing these modes. For the striped wind we have:
\begin{eqnarray}
\mu &=& \Gamma (1+\sigma) \label{muwind} \\
\nu &=& \left[\Gamma^2(1+\sigma)-(1+\sigma/2)\right](\Gamma^2-1)^{-1/2} \\
\eta &=& b^2\Gamma \sigma \beta \label{etawind}
\end{eqnarray}where $\sigma$ is the wind magnetization and $b$ is the normalized DC component of the
magnetic field. This quantity is zero at the equatorial plane, where regions of alternating field are equal 
in width and cancel out in the phase average, and rises to its maximum value, $b=1$, at latitude equal to the 
inclination angle $\alpha$, where the alternating field component disappears.
The same quantities $\mu$, $\nu$  and $\eta$ can be calculated for the superluminal wave. Since our treatment of these 
modes took
place in the H-frame, we calculate the quantities $\left< J \right>$, $\left< T^{00}\right>$, $\left< T^{01}\right>$, 
$\left< T^{11} \right>$ and $\left< E\right>^2/(4\pi)$ and transform them to the laboratory frame using the velocity 
$c^2/v_{\phi}$, as explained above. 

\begin{figure}[pt]
\centerline{\psfig{file=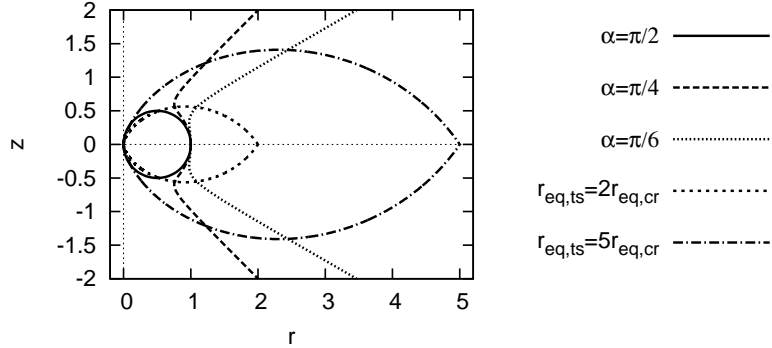,width=11cm}}
\vspace*{8pt}
\caption{The critical surfaces for three values of the inclination angle $\alpha$, along with two example shock surface, normalized to equatorial radius $\rho_{\rm eq, ts}$ two and five times larger than the equatorial critical radius $\rho_{\rm eq, 
cr}$. The pulsar is at the point $(0,0)$ and one half of the poloidal plane is shown. The variable $r$ is normalized to the
equatorial value of $R$.
\label{f2}}
\end{figure}

Equating Eqns.~(\ref{mugeneral})-(\ref{etageneral}) calculated for the superluminal mode to Eqns.~
(\ref{muwind})-(\ref{etawind}), one can find "jump conditions" for the transition from the striped wind to the superluminal 
mode, for certain values of $\Gamma$, 
$\sigma$ and latitude, or equivalently $b$. We can then re-introduce the conservation of particle flux into our system of
equations, in order to find the radial dependence of these jump conditions\cite{arkakirk}.

\section{Results and discussion}

In Fig.~\ref{f1} we have plotted as a function of normalized radius $R=\rho \mu/a_L$ the inverse hyperbolic tangent of the 
refractive index $c/v_{\phi}$ of the wave into which a striped wind with $\Gamma = 100$ and $\sigma = 100$ can convert.
The quantity $a_L$ is a dimensionless parameter given by: 
\begin{equation}
a_{\rm L}  = \sqrt{\frac{4\pi e^2}{m^2c^5} \frac{dL}{d\Omega}} = a_{\rm L0} \sin\theta,
\end{equation}where by $dL/d\Omega$ the distribution of the pulsar's luminosity in solid angle $\Omega$ is meant and 
$\theta$ is the colatitude (i.e. the polar angle). We have assumed that the distribution of the pulsar's luminosity follows the 
prescription 
\begin{equation}
\frac{dL}{d\Omega} = L_0 \sin^2\theta
\end{equation}and separated the angular dependence from $a_{\rm L}$.

As we can see in Fig.~\ref{f1}, there is a minimum normalized distance $R_{\rm cr}$ outside which superluminal waves can 
propagate. For $b=0$, $R_{\rm cr} \simeq 1$, but it rises as one goes to higher latitudes. On the plot on the right of 
Fig.~\ref{f1} the phase averaged Lorentz factor of the particles, $\left< \gamma' 
\right>$ is shown for the same values of $b$. From this plot it is seen that during the conversion from one wave to another, \
energy is transferred from fields to particles. In the striped wind the Lorentz factor of each particle is equal to the bulk 
Lorentz factor of the outflow, which
in the plotted example is $\Gamma=100 \ll \mu$. However, in the superluminal wave, for low values of $b$, $\left< \gamma' 
\right> \simeq \mu$, which means that almost all the energy of the outflow is now carried by particles. Even for values of $b$ 
close to unity, 
$\left< \gamma' \right> \gg \Gamma$, meaning that the particles get accelerated in the whole wind, and the ratio of the
Poynting to the kinetic energy flux decreases during the transition from the subluminal to the superluminal regime.

In 
Fig.~\ref{f2}, the geometrical critical surface is shown (radius normalized as $\rho \rightarrow \rho \mu/a_{\rm L0}$), outside 
which superluminal waves can propagate, for different values 
of the inclination $\alpha$.
Superimposed on the critical surfaces is the example of a shock surface, the shape of which is taken from 
Ref.~\refcite{lyubarsky2}. The normalization of the shock radius in comparison to the critical radius in Fig.~\ref{f2} is arbitrary. 
Generally it depends on
the individual pulsar in question. For isolated pulsars like the Crab and the Vela, the shock (located at $\sim 
10^9r_{\rm LC}$ and $\sim 10^8 r_{\rm LC}$ respectively\cite{observers}) is much farther out 
than the critical surface (located at $\sim 10^7 r_{\rm LC}$ and $\sim 10^6r_{\rm LC}$ respectively) for practically the whole of 
the wind. For pulsars in binary systems, though, where the shock is 
pushed closer to the pulsar by the outflow from a companion star\cite{pulsarbinary}, it is possible that the critical surface is partly or entirely 
outside the shock, so that no superluminal waves can propagate. In highly eccentric systems, the possibility arises that the 
termination shock migrates during one orbit from a location inside to one outside the critical surface, so that the
characteristics of the wind just before the termination shock are orbitally modulated during one period of the system.

We conclude that the conversion of the striped wind to a superluminal mode is possible in many objects, especially in
isolated pulsars, and it is accompanied by an acceleration of the particles of the outflow to $\left< \gamma' \right> \gg 
\Gamma$, which might influence the physics of particle acceleration and the structure of the termination shock.


\end{document}